\documentstyle[12pt]{article}
\newcommand{\tr}{\mathop{\rm tr}\nolimits}
\normalbaselines
\catcode `\@=11
\@addtoreset{equation}{section}

\def\section{\@startsection {section}{1}{\z@}{-3.5ex plus -1ex minus
     -.2ex}{2.3ex plus .2ex}{\normalsize\bf}}
\def\subsection{\@startsection{subsection}{2}{\z@}{-3.25ex plus -1ex minus
-.2ex}{1.5ex plus .2ex}{\normalsize\bf}}

\def\thebibliography#1{\section*{References}
\list
  {[\arabic{enumi}]}{\settowidth\labelwidth{[#1]}\leftmargin\labelwidth
  \advance\leftmargin\labelsep
  \usecounter{enumi}}
  \def\newblock{\hskip .11em plus .33em minus -.07em}
  \sloppy
  \sfcode`\.=1000\relax}

\catcode `\@=12

\begin{document}

\begin{titlepage}
\vspace*{2.5cm}

\begin{center}

{\bf To the problem on the Universe evolution and
  the condensed description in the field theory}\vspace{0.3cm}\\

\medskip

{\bf V.M. Koryukin}
\vspace{0.3cm}\\

 Mari State Technical University,\\
 sq. Lenin 3, Yoshkar--Ola, 424024, Russia\\
 e-mail: koryukin@marstu.mari.ru

\end{center}

\vspace*{0.5cm}

\noindent
 On the basis of hypotheses, 
 that a density of weakly interacting particles in the Universe
 has an order of nuclear matter density or more
 the Lagrangian is offered,
 through which one can be obtained
 a propagator of a vector boson with a non-zero rest-mass.
 The dependence of vector bosons masses on the time allows
 to explain the availability of the hot stage
 of the Universe evolution,
 not using to the hypothesis of the Universe expansion.



\end{titlepage}

 From last astronomical observations (see for example~\cite{lei}) it follows
 that not more than $5\%$ of the Universe matter has the baryon nature.
 Thereof it is convenient to divide the Universe matter
 on rapid and slow subsystems. We shall consider
 that all known particles (it is possible excluding only neutrinos)
 belong to the rapid subsystem.
 Thus we ought to consider fundamental particles
 as coherent frames in open systems
 characterized by quasi--group structure.
 It results in necessity of usage inhomogeneous
 (quasi--homogeneous) space-time
 endowed by a nontrivial geometrical structure.

 Let's consider the Boltzmann hypothesis of the Universe birth
 owing to a gigantic fluctuation not in an empty space but in a medium
 which consists of weakly interacting particles
 characterized by zero temperature and forming the Bose condensate.
 Certainly,
 if the particles are fermions
 they should be in the coupled state.
 For the description of such state of the Universe matter
 (this state we shall consider pure one)
 it is necessary to introduce
 an amplitude of probability ${\cal B}$ with components
 ${\cal B}_a^b(\omega) \;$
 ($a, b, c, d, e, f, g, h = 1, 2..., r $)
 dependent from points $\omega$ of a manifold $M_r$
 (not excepting a limiting case $r \to \infty $).
 In this case we can not define the metric
 but for its definition we need a density matrix
 $\rho({\cal B})$
 (the rank of which equals to $1$ for a pure state),
 determining its standard mode ${\cal B}{\cal B}^+ =
 \rho\tr({\cal B}{\cal B}^+)$
 ($\tr\rho=1$, $\rho^+ =\rho$,
 the top index $``+''$ is the symbol
 of the Hermitian conjugation).

 Let as a result of a fluctuation the disintegration
 of the Bose condensate will begin with formation of fermions
 (for their description we shall introduce
 an amplitude of probability $\Psi $)
 and with an increase of pressure in some local area of Universe
 (in addition some time the temperature of the Universe particles
 could remain equal or close to zero point
 --- so--called the inflation period).
 As a result the rank of the density matrix $\rho $ will begin to grow
 that characterizes the appearance of mixed states.
 An inverse process of relaxation
 (characterized by the formation of
 the Bose condensate and by the pressure decline)
 should go with an energy release
 which will go on heating of the Fermi liquid
 with formation of excited states --- of known charged fermions
 (quarks and leptons).
 From this moment it is possible to introduce the metric
 and use the results obtained for the hot model of Universe
 (with those by inflationary modifications
 which have appeared recently~\cite{kla})
 interpreting the Universe evolution as the process
 characterized by the increase of the entropy $S = - \tr(\rho\ln\rho)$.
 Now Universe is at that stage of an evolution
 when the dominating number of particles has returned
 to the Bose condensate state.
 They are also displaed only at a weak interaction
 with particles of a visible matter.

 For the description of a matter
 it is convenient to use differentiable fields
 given in a differentiable manifold $M_n$
 which one we shall call as a space-time
 and the points $x$ of which one will have coordinates $x^i$
 ($i, j, k, l... = 1, 2..., n$).
 Probably the rank of the density matrix $\rho$ equals $n$,
 but it is impossible to eliminate
 that the generally given equality is satisfied only approximately
 when some components of a density matrix can be neglected.
 In any case we shall consider that among fields ${\cal B}$
 the mixtures $\Pi_a^i$ were formed with non-zero vacuum means $h_a^i$
 which determine differentiable vector fields $\xi_a^i(x)$ as:
\begin {equation}
\label{1}
 \Pi^i_a = {\cal B}_a^b~\xi_b^i
\end {equation}
 for considered area $\Omega_n\subset M_n$
 (field $\xi_a^i(x)$ determine a differential of a projection
 $d\pi$ from $\Omega_r \subset M_r$ in $\Omega_n$).
 It allows to define a space--time $M_n$
 as a Riemannian manifold,
 the basic tensor $g_{ij}(x)$ of which we shall introduce
 through a reduced density matrix $\rho'(x)$.
 
 So let components $\rho_i^j$ of a reduced density
 matrix $\rho'(x)$ are determined by the way:
\begin{equation}
\label{2}
 \rho_i^j = \xi^+{}_i^a~\rho_a^b~\xi_b^j~/
 ~(\xi^+{}_k^c \rho_c^d \xi_d^k) .
\end{equation}
 and let fields
\begin{equation}
\label{3}
 g^{ij} = \eta^{k(i}\rho_k^{j)}~(g^{lm}\eta_{lm})
\end{equation}
 are components of a tensor of a converse to the basic tensor
 of the space--time $M_n$.
 By this components $g_{ij}(x)$ of the basic tensor
 must be the solutions of following equations:
 $g^{ij} g_{ik} = \delta^j_k$.
 (Hereinafter $\eta_{ij}$ are metric tensor components
 of a tangent space to $M_n$
 and $\eta^{ik}$ are determined as the solution of equations:
 $\eta^{ij} \eta_{ik} = \delta^j_k $.)

 The influencing of the macroscopic observer will display
 in an approximation of the transition operator $T$
 by the differential operators $\partial_i$.
 First of all we shall require
 that for fermion fields deviations
 $X_a(\Psi) = T_a(\Psi)-\xi_a^i\partial_i\Psi$
 were minimum in the ``mean''~\cite{k1}.
 For this purpose we shall consider a following integral
\begin{equation}
 \label{4}
 {\cal A} = \int\limits_{\Omega_n} {\cal L}(\Psi) d_nV =
 \int\limits_{\Omega_n}\kappa~\overline{X}^a(\Psi)~
 \rho_a^b(x)~X_b(\Psi)~d_nV
\end{equation}
 ($\kappa$ is a constant, ${\cal L}(\Psi)$ is a Lagrangian;
 the bar means the Dirac conjugation
 that is to be the superposition of Hermitian conjugation and
 the spatial inversion)
 which is the action.
 Let the action ${\cal A}$ is quasi-invariant
 at infinitesimal substitutions
 $x^i \to x^i+\delta x^i=x^i+\delta\omega^a\xi_a^i(x), \;
 \Psi \to \Psi+\delta\Psi=\Psi+\delta\omega^aT_a(\Psi)$
 of Lie local loop $G_r$,
 the structural tensor components $C_{ab}^c$ of which 
 satisfy to Jacobi generalized identity~\cite{k1}.
 The given requirement causes to introduce the full
 Lagrangian (instead of a Lagrangian
 ${\cal L}(\Psi)$) recording it as
\begin{equation}
\label{5}
 {\cal L}_t = {\cal L}(\Psi) +
 \kappa'_o {\cal F}_{ab}^c {\cal F}_{de}^f
 [t^{ad} (s_c^e s_f^b - \upsilon s_c^b s_f^e) +
 t^{be} (s^a_f s_c^d - \upsilon s^a_c s^d_f) +
 u_{cf} (t^{ad} t^{be} - \upsilon t^{ab} t^{de})]/4
\end{equation}
 ($\kappa'_o, \upsilon$ are constants).
 In addition  intensities ${\cal F}_{ab}^c(B)$ of
 the boson (gauge) fields ${\cal B}_a^c(x)$ will look like
\begin{equation}
\label{6}
 {\cal F}_{ab}^c = \Theta_d^c~(\Pi_a^i~\partial_i {\cal B}_b^d -
 \Pi_b^i~\partial_i {\cal B}_a^d + \Xi_{ab}^d) ,
\end{equation}
 where
\begin{equation}
\label{7}
 \Theta_b^c = \delta_b^c - \xi_b^i~\Pi_i^d~({\cal B}_d^c -
 \beta_d^c) , \quad
 \Xi_{ad}^b = ({\cal B}_a^c~T_c{}^e_d
 - {\cal B}_d^c~T_c{}^e_a)~{\cal B}_e^b -
 {\cal B}_a^c~{\cal B}_d^e~C_{ce}^b .
\end{equation}
 Hereinafter a selection of fields $\Pi_i^a(å)$ and $\beta_c^a$
 are limited by the relations:
\begin{equation}
\label{8}
 \Pi_j^a~\Pi_a^i = \delta_j^i , \quad
 \beta_c^a~\xi_a^i = h_c^i
\end{equation}
 ($\delta_j^i$ are Kronecker deltas).
 If
\begin{equation}
\label{9}
 s_a^b = \delta_a^b, \quad
 t^{ab} = \eta^{ab}, \quad
 u_{ab} = \eta_{ab}
\end{equation}
 ($\eta_{ab}$ are metric tensor components of the flat space
 and $\eta^{ab}$ are tensor components of a converse to basic one)
 then the given Lagrangian is most suitable one at the description
 of the hot stage of the Universe evolution
 because it is most symmetrical one
 concerning intensities of the gauge fields ${\cal F}_{ab}^c$.
 What is more we shall require the realization of the correlations:
 $T_a{}_á^b~\eta^{cd} + T_a{}_c^d~\eta^{cb} = 0$,
 that the transition operators
 $T_a{}_á^b$ generate the symmetry, which follows from
 the made assumptions.
 In absence of fields $\Pi_a^i(x)$ and $\Psi(x)$ at earlier stage
 of the Universe evolution
 the Lagrangian (5) becomes even more
 symmetrical (${\cal L}_t\propto {\cal B}^4$),
 so that the formation of fermions
 (the appearance of fields $\Psi$
 in a full Lagrangian ${\cal L}_t$)
 from primary bosons is a necessary condition (though 
 not a sufficient one) of the transition of Universe
 to the modern stage of its development with a spontaneous
 symmetry breaking.
 Only the formation of the Bose condensate
 from pairs of some class of
 fermions (the neutrinos of different flavors)
 has resulted in a noticeable growth of
 rest--masses of those vector bosons ($W^+, W^-, Z^o$),
 which interact with this class of fermions.
 In parallel there could be a growth of rest--masses
 and other fundamental particles,
 though and not all
 (photon, directly with a neutrino not interacting,
 has not a rest-mass).

 Let's connect non-zero vacuum means $\beta^b_a $ of gauge
 fields $B^b_a$ with a spontaneous violation of a symmetry,
 which has taken place in the early Universe
 and which is a phase transition
 with a formation of Bose condensate from fermion pairs.
 The transition to the modern stage of the Universe evolution
 for which it is possible to suspect the presence of cluster states
 of weakly interacting particles
 will be expressed in following formula for
 tensors $s_a^b$, $t^{ab}$, $u_{ab}$ and $h_i^a$:
$$
 s_a^b = s~\xi_a^i~h^b_i +
 \xi_a^{\underline{c}}~\epsilon_{\underline{c}}^b, \quad
 t^{ab} = t~\epsilon^a_{(i)}~\epsilon^b_{(j)}~\eta^{(i)(j)} +
 \epsilon^a_{\underline{c}}~\epsilon^b_{\underline{d}}~
 \eta^{\underline{cd}},
$$
\begin{equation}
 \label{10}
 u_{ab} = u~\xi_a^i~\xi_b^j~h_i^c~h_j^d~\eta_{cd} +
 \xi_a^{\underline{c}}~\xi_b^{\underline{d}}~
 \eta_{\underline{cd}}, \quad
 h_i^a = h_i^{(j)}~\epsilon_{(j)}^a
\end{equation}
 ($(i), (j), (k), (l), ... = 1, 2, ..., n;\; \underline{a},
 \underline{b}, \underline{c}, \underline{d}, \underline{e}
 = n+1, n+2, ..., n+\underline{r};\; \underline{r} /r \ll 1$),
 where fields $h^{(j)}_i(x)$, taking into account the relations (10),
 are uniquely determinated from equations: $h^a_k~h_a^i = \delta_k^i $.
 Similarly tensors $\eta^{(i)(j)}, \eta^{\underline{ab}}$
 are determined from equations:
 $\eta^{(i)(k)}\eta_{(j)(k)} = \delta_{(j)}^{(i)},
 \eta^{\underline{ab}}\eta_{\underline{cb}}=
 \delta_{\underline{c}}^{\underline{a}}$,
 while tensors $\eta_{(i)(j)}, \eta_{\underline{ab}}$
 are determined as follows
 $\eta_{(i)(k)}=\eta_{ab}~\epsilon^a_{(i)}~\epsilon^b_{(k)}, \;
 \eta_{\underline{ab}}=\eta_{cd}~
 \epsilon_{\underline{a}}^c~\epsilon_{\underline{b}}^d$.
 We shall connect constants $\epsilon^a_{(i)}$, $\epsilon^a_{\underline{b}}$
 with a selection of the gauge fields $\Pi_i^a(x)$
 recording them by in the form
\begin{equation}
 \label{11}
 \Pi_i^a = \Phi_i^{(j)}~\epsilon_{(j)}^a +
 \Phi_i^{\underline{b}}~\epsilon_{\underline{b}}^a
\end{equation}
 and let $\epsilon_{\underline{b}}^a = 0$.
 Besides we shall apply the decomposition of fields $B_b^a(x)$
 in the form
\begin{equation}
\label{12}
 B_c^a=\zeta_i^a~\Pi_c^i
 + \zeta_{\underline{b}}^a~A_c^{\underline{b}} ,
\end{equation}
     where
 $A_b^{\underline{a}}=B_b^c~\xi_c^{\underline{a}}$.
 In addition components of intermediate tensor fields
 $\xi_a^i (x)$, $\xi_a^{\underline{b}}(x)$,
 $\zeta_i^a(x)$, $\zeta_{\underline{b}}^a(x)$
 should be connected by the relations:
 $\zeta_i^a\xi_a^j = \delta_i^j,\;
 \zeta_i^a\xi_a^{\underline{b}} = 0, \;
 \zeta_{\underline{b}}^a\xi_a^j = 0, \;
 \xi_a^{\underline{c}}\zeta^a_{\underline{b}}
 = \delta_{\underline{b}}^{\underline{c}}$.

 Let $n = 4$, $\; \upsilon = 2$, $\; t u = s^2$, $\;
 T_c{}_{(k)}^a=T_c{}_b^a\epsilon^b_{(k)}
 =T_c{}_{(k)}^{(i)}\epsilon_{(i)}^a$, $\;
 T_i{}_{(j)}^{(k)}=\zeta_i^a T_a{}_{(j)}^{(k)}$, $\;
 T_{\underline{b}}{}_{(j)}^{(k)}=
 \zeta_{\underline{b}}^a T_a{}_{(j)}^{(k)}$ and
\begin{equation}
 \label{13}
 T_a{}_{(k)}^{(i)}~\eta^{(k)(j)}+T_a{}_{(k)}^{(j)}~\eta^{(k)(i)} =0 ,
\end{equation}
 so that the full Lagrangian (5) will be rewritten as follows
$$
 {\cal L}_t = {\cal L}(\Psi, D\Psi) + 
 \eta^{(j)(m)}~[\kappa_o~E_{(i) (j)}^{\underline{a}}
 ~E_{(k) (m)}^{\underline{b}}~\eta^{(i) (k)}
 ~\eta_{\underline{a}\underline{b}} +
$$
\begin{equation}
\label{14}
 \kappa_1~(F_{(i) (j)}^{(k)}
 ~F_{(l) (m)}^{(n)}~\eta^{(i) (l)}~\eta_{(k) (n)} +
 2~F_{(i)(j)}^{(k)}~F_{(k) (m)}^{(i)} -
 4~F_{(i) (j)}^{(i)}~F_{(k) (m)}^{(k)})]/4 ,
\end{equation}
 where
\begin{equation}
\label{15}
 \kappa_o = \kappa'_o~t^2, \quad \kappa_1 = \kappa'_o~t~s^2 .
\end{equation}
$$
 F_{(i)(j)}^{(k)}=F_{ab}^á
 ~\epsilon^a_{(i)}~\epsilon^b_{(j)}
 ~h_c^l~h_l^{(k)}=
 \Phi^{(k)}_l~F^l_{mn}~\Phi^m_{(i)}~\Phi^n_{(j)}=
$$
\begin{equation}
\label{16}
 (\Phi{(i)}^l\nabla_l\Phi_{(j)}^m-
 \Phi_{(j)}^l\nabla_l\Phi_{(i)}^m)~\Phi_m^{(k)}+
 \Phi_{(i)}^lT_l{}_{(j)}^{(k)}-\Phi_{(j)}^lT_l{}_{(i)}^{(k)} +
 A_{(i)}^{\underline{a}}~T_{\underline{a}}{}_{(j)}^{(k)}
 -A_{(j)}^{\underline{a}}~T_{\underline{a}}{}_{(i)}^{(k)},
\end{equation}
$$
 E_{ij}^{\underline{a}}=
 E_{(k)(l)}^{\underline{a}}~\Phi^{(k)}_i~\Phi^{(l)}_j
 =F_{bc}^d~\xi_d^{\underline{a}}
 ~\epsilon^b_{(k)}
 ~\epsilon^c_{(l)}~\Phi^{(k)}_i~\Phi^{(l)}_j=
$$
\begin{equation}
\label{17}
 \nabla_i A_j^{\underline{a}} - \nabla_j A_i^{\underline{a}}+
 A_i^{\underline{b}}~A_j^{\underline{c}}
 ~C_{\underline{b}\underline{c}}^{\underline{a}}+
 C_{i\underline{b}}^{\underline{a}}~A_j^{\underline{b}}
 -C_{j\underline{b}}^{\underline{a}}~A_i^{\underline{b}}
 +C_{ij}^{\underline{a}} ,
\end{equation}
\begin{equation}
\label{18}
 \Phi_{(k)}^i = \Pi_a^i~\epsilon^a_{(k)},\quad
 A_i^{\underline{b}}=A_{(j)}^{\underline{b}}~\Phi^{(j)}_i,
\end{equation}
\begin{equation}
 \label{19}
 C_{\underline{a}\underline{b}}^{\underline{c}}=
 \zeta_{\underline{a}}^e\zeta_{\underline{b}}^d
 C_{ed}^g~\xi_e^{\underline{c}},\;
 C_{i\underline{a}}^{\underline{c}}=
 (\zeta_i^b\zeta_{\underline{a}}^d C_{bd}^e+
 \partial_i\zeta_{\underline{a}}^e)\xi_e^{\underline{c}},\;
 C_{ij}^{\underline{c}}=(\zeta_i^a\zeta_j^b C_{ab}^d+
 \nabla_i\zeta_j^d-\nabla_j\zeta_i^d)\xi_d^{\underline{c}} .
\end{equation}
 As a result of an equation of fields $\Phi^i_{(j)}(x)$
 may be received in a standard manner~\cite{tre}
 as the Einstein gravitational equations
$$
 D_i\Psi~\frac{\partial{\cal L}}{\partial D_j\Psi}~g_{jk} -
 g_{ik}~{\cal L}(\Psi, D_m\Psi) +
 \kappa_o~\eta_{\underline{a}\underline{b}}~g^{jl}
 ~(E^{\underline{a}}_{ij}~E_{kl}^{\underline{b}} -
 \frac14~g_{ik}~g^{mn}~E^{\underline{a}}_{jm}
 ~E_{ln}^{\underline{b}}) 
$$
\begin{equation}
\label{20}
 = \kappa_1~(2~R_{jik}{}^j - g_{ik}~g^{lm}~R_{jlm}{}^j)
\end{equation}
 ($R_{ijk}{}^l$ is the curvature tensor
 of the connection $\Gamma_{ij}^k$ of
 Riemannian space--time $M_n$;
 $\kappa_o = 1/(4\pi)$,
 $\kappa_1 = 1/(4\pi G_N)$, $G_N$ is the gravitational constant;
 hereinafter the system of units are used $h/(2\pi) =c=1$,
 where $h$ is the Planck constant, $c$ is the speed of light).
 Naturally,
 that the Einstein equations express the modern physical state
 of the Universe matter.
 All this confirms a capability for
 interpretations of fields $\Phi_{(j)}^i(x)$ or
 fields $\Phi^{(j)}_i(x)$
 as gravity potentials,
 but taking into account their dependence from properties of medium
 (vacuum),
 and also the opinion
 to consider components $g_{ij}(x)$ of a metric tensor
 of the space--time by potentials of a gravitational field,
 it is meaningful to call $\Phi_{(j)}^i(x)$ and $\Phi^{(j)}_i(x)$
 by polarization fields.
 It is necessary to note
 that the condensed description of a medium, consisting of
 weakly interacting particles,
 by polarization (gravitational) fields $\Phi_i^{(k)}(x)$
 allows to ``hide'' the Bose condensate with
 the help of a nontrivial geometrical structure,
 in particular,
 using a Riemannian space--time of General Relativity.

 Let's study an approaching,
 in which the space--time is possible
 to consider as a Minkowski space,
 the fields $\Phi_i^{(k)}$, $\Phi^i_{(k)}$ are constants
 and let $\underline{r} =1$,
 that assumes
 $C_{\underline{a}\underline{b}}^{\underline{c}}=0$.
 For obtaining equations of fields $A_i^{\underline{b}}(x)$ in
 Feynman perturbation theory the calibration should be fixed.
 For this we shall add the following addend:
\begin{equation}
\label{21}
 {\cal L}_q = \kappa_o~q_{\underline{b}\underline{b}}~
 g^{ij}~g^{kl}~(\partial_i A_j^{\underline{b}}
 - q_o~C_i~A_j^{\underline{b}})~(\partial_k A_l^{\underline{b}}
 - q_o~C_k~A_l^{\underline{b}}) /2
\end{equation}
 to a Lagrangian (14), where
 $q_o = \eta_{\underline{b}\underline{b}}
 / q_{\underline{b}\underline{b}}, \;
 C_i = C_i{}_{\underline{b}}^{\underline{b}}.\;$
 Besides let
\begin{equation}
\label{22}
 T_a{}_{(k)}^{(i)}~\eta^{(j)(k)}+
 T_a{}_{(k)}^{(j)}~\eta^{(i)(k)}
 = \epsilon_a^{\underline{b}}~t_{\underline{b}}~\eta^{(i)(j)} .
\end{equation}
 As a result of this equations of a vector field $A_i^{\underline{b}}(x)$
 will be written as:
\begin{equation}
\label{23}
 g^{jk} [\partial_j\partial_k A_i^{\underline{a}} -
 (1 - 1/q_o)\partial_i\partial_j A_k^{\underline{a}} +
 (1 - q_o) C_i C_j A_k^{\underline{a}} ] +
 m^2 A_i^{\underline{a}} = I_i^{\underline{a}} /\kappa_o ,
\end{equation}
 where $I_i^{\underline{a}} =
 \frac{g_{ij}}{\eta_{\underline{a}\underline{a}}}~
 \frac{\partial {\cal L}(\Psi)}{\partial A_j^{\underline{a}}}$ and
\begin{equation}
\label{24}
 m^2 = (n-1)(n-2)\kappa_1 t^2_{\underline{a}}/
 (2\kappa_o \eta_{\underline{a}\underline{a}})
 - g^{jk} C_j C_k .
\end{equation}
 Notice that owing to the vacuum polarization
 ($C_i\ne 0 $) the propagator of a vector boson has
 the rather cumbersome view~\cite{k2}
$$
 D_{ij}(p)=[(1-q_o)\frac{(p_ip_j-C_iC_j)(p^kp_k-q_om^2)
 +(1-q_o)p^kC_k(p_iC_j+C_ip_j)}{(p^lp_l-q_om^2)^2+
 (1-q_o)^2(p^lC_l)^2} 
$$
\begin{equation}
 \label{25}
 - g_{ij}]/(p^m p_m-m^2) ,
\end{equation}
 which is simplified and receives the familiar form
 ($ - g_{ij}/(p_k p^k - m^2)$, $p^k$
 is the 4-momentum,
 and $m$ is the mass of the vector boson)
 only in the Feynman calibration ($q_o=1 $).

 So, the transition to the hot state of Universe
 was connected
 with the destruction of the Bose condensate
 and the increase of the Fermi gas pressure accordingly.
 In addition some time a temperature of background particles
 of Universe could remain equal or close to zero
 (the stage of the inflation).
 As a result the rest--mass of $W^+, W^-, Z^o$ bosons
 have decreased so, that the weak interaction
 has stopped to be weak and all (or nearly so all)
 particles from a ground (vacuum) state started to participate in
 an installation of a thermodynamic equilibrium.
 The given phenomenon also has become the cause of an
 apparent increase of a density of particles in the early
 Universe. Suggesting,
 that mean density $n_o$ of particles in the Universe did not vary
 at the same time
 and the script of the hot model
 in general is correct,
 we come to its following estimation
 $n_o\sim m^3_{\pi} \sim 10^{-3} GeV^3$
 ($m_{\pi}$ is a mass of a $\pi$ meson).
 This result allows to give explanation to a known ratio~\cite{wei}
 $H_o/G_N\approx m^3_{\pi}$,
 if to consider,
 that the Hubble constant $H_o$ gives
 an estimation $1/H_o$ to the length
 $l\sim 1/(n_o\sigma_{\nu})$
 of a free run of particle in ``vacuum'' on the modern
 stages of the Universe evolution
 ($\sigma_{\nu}$ is a scattering cross--section
 of neutrinos on a charged particle)
 and to take into account the estimation
 given earlier~\cite{k3} for the gravitational constant $G_N$
 ($G_N\sim\sigma_{\nu}$).
 Thus the gravitational constant $G_N$
 is inversely proportional to the time of a free run
 of a charged particle in the neutrinos medium
 characterizing a kinetic phase of a relaxation process
 in the Universe.

\end{document}